\documentclass[letter, traditabstract]{aa}
\usepackage{graphicx}
\usepackage{epsfig}
\usepackage{color}
\usepackage{txfonts}
\usepackage{natbib}
\bibpunct{(}{)}{;}{a}{}{,} 

\begin{document}

\title{The formation of a thick disk through the heating of a thin disk:\\
 Agreement with orbital eccentricities of stars in the solar neighborhood}
\titlerunning{Orbital eccentricities compared with results of numerical simulations}

\author{Paola Di Matteo\inst{1}, Matthew
  Lehnert\inst{1},  Yan Qu\inst{1}, \and Wim van Driel\inst{1}}

\authorrunning{Di Matteo et al.}

\institute{GEPI, Observatoire de Paris, CNRS, Universit\'e
  Paris Diderot, 5 place Jules Janssen, 92190 Meudon, France\\
\email{paola.dimatteo@obspm.fr}
}

\date{Accepted, Received}

\abstract{We study the distribution of orbital eccentricities of stars in
thick disks generated by the heating of a pre-existing thin stellar disk
through a minor merger (mass ratio 1:10), using N-body/SPH numerical
simulations of interactions that span a range of gas fractions in
the primary disk and initial orbital configurations. The resulting
eccentricity distributions have an approximately triangular shape,
with a peak at 0.2-0.35, and a relatively smooth decline  towards
higher values.  Stars originally in the satellite galaxy tend to have
higher eccentricities (on average from $e=0.45$ to $e=0.75$), which is in
general agreement with the models of Sales and collaborators, although in
detail we find fewer stars with extreme values and no evidence of their secondary peak around $e=0.8$. 
The absence of this high-eccentricity feature results in a distribution  that qualitatively matches the observations. Moreover, the
increase in the orbital eccentricities of stars in the solar neighborhood
with vertical distance from the Galactic mid-plane recently found by
Diericxk and collaborators can be qualitatively reproduced by our models, but
only if the satellite is accreted onto a direct orbit. We thus speculate
that if minor mergers were the dominant means of formating
the Milky Way thick disk, the primary mechanism should be
merging with satellite(s) on direct orbits.}


\maketitle

\section{Introduction}

\citet{sales09} proposed using the orbital eccentricity
of stars in a thick disk as a way of constraining the main physical
mechanism responsible for its formation. They pointed out the distinctive
differences between the eccentricity distributions predicted for a
comprehensive range of formation mechanisms for the thick disk; the four
mechanisms they considered were: the heating of a pre-existing thin disk
by a varying gravitational potential in the thin disk, which causes radial
migration and a thickening of the stellar disk \citep{roskar08}; heating
of a pre-existing thin disk by minor mergers  \citep{villalobos08};
accretion of disrupted satellites  \citep{abadi03}; and formation
by gas-rich mergers \citep{brook04, brook05}. Their results can be
summarized as: (1) the accretion scenario predicts a broad distribution
of stellar eccentricities, $e$, which are quite symmetric with respect to its
peak at around $e=0.5$; (2) the radial migration scenario produces a
thick disk whose stars have eccentricities peaking around $e=0.25-0.3$,
and whose distribution is quite narrow and symmetric around its peak;
(3) the heating of a pre-existing thin disk by a 1:5 mass ratio merger
produces a distribution with a peak around $e=0.25$, with a secondary peak
and tail at high eccentricities composed mostly of stars originally in the
satellite galaxy; and (4) gas-rich mergers produce a result quite similar
to the previous scenario, except for the absence of the secondary peak.
The strength of this work is that it considers a number of scenarios for
the formation of the thick disk -- each of which apparently produces a
unique (and testable) signature.

A comparison of the eccentricity distributions predicted by these
four models with that of stars populating the thick disk in the solar neighborhood led \citet{dierickx10} 
 to reject  the above-mentioned models (1), (2) and (3), which leads them to
propose gas-rich mergers at early epochs as the most probable mechanism for
the formation of the Galactic thick disk.  Similarly,  \citet{wilson10} found that the eccentricity distribution of a sample of solar neighborhood thick disk stars from RAVE survey data appears to be most consistent with the gas-rich merger scenario.
  Within the limitations of any of these comparisons, their conclusions appear well justified.
However, the question arises of whether this is the complete story.

To make this comparison based on a comprehensive analysis of the
effects of minor mergers, a wide range of initial orbital parameters
needs to be considered in numerical simulations.  The aim of this Letter
is to compare the observed distribution of the eccentricities of thick
disk stars in the solar neighborhood with those predicted by twelve
1:10 minor mergers simulations, with and without gas in the primary,
i.e. the most massive galaxy, and in the satellite, that have a range
of initial orbital characteristics.
 
\section{Models}\label{model}

The simulations we analyzed  are part of a set of minor
merger simulations with a mass ratio of 1:10, described extensively in
\citet{chili10}. The primary and the satellite galaxy each consist of
a spherical non-rotating halo and a central bulge, both modelled as Plummer
spheres, and a stellar and an optional gas disk represented by Myamoto-Nagai
density profiles. Both the primary and the satellite galaxies span a
range of morphologies, with bulge-to-disk mass ratios of between 0.2 and 0.25 and
gas-to-stellar disk fractions, $f_{gas}$, from 0 to 0.2.  We analyzed twelve of the encounters labelled  gS0dS0, gSadSa, and gSbdSb in \citet{chili10} containing, respectively,  $f_{gas}$=0, 0.1, and 0.2.
A total of $N_{\rm TOT}=528~000$ particles were used in all simulations, distributed between the primary galaxy and the satellite. We  tested the dependence of the results on the number of particles used by running an additional simulation with 2.5 times more particles, $N_{\rm TOT}=1~320~000$.
After the particles in
each simulation had relaxed, the two galaxies were placed at an initial
distance of 100 kpc, with a variety of initial relative velocities (id=01dir, 01ret, 02dir, 02ret in Table 9 of \citet{chili10}), to simulate
different orbits. For a complete description of the initial galaxy models,
the number of particles employed, and the orbital characteristics, we refer the reader
to Tables 1, 3, 6, and 9 in \citet{chili10}.

All simulations were run using the Tree-SPH code described in
\citet{seme02}. The gas is assumed to be isothermal at a temperature
of 10$^4$ K. Prescriptions for star formation and feedback from
supernovae explosions are also included. The equations of motion are
integrated using a leap-frog algorithm, with a fixed time step of 0.5
Myr. For the evaluation of the gravitational forces, a softening length
$\epsilon$=200 pc (or $\epsilon$=150 pc for the high-resolution simulation) is employed. With these choices, the relative error
in the conservation of the total energy is about $10^{-6}$ per time step.

\begin{figure*}
\centering
\includegraphics[width=4.7cm,angle=270]{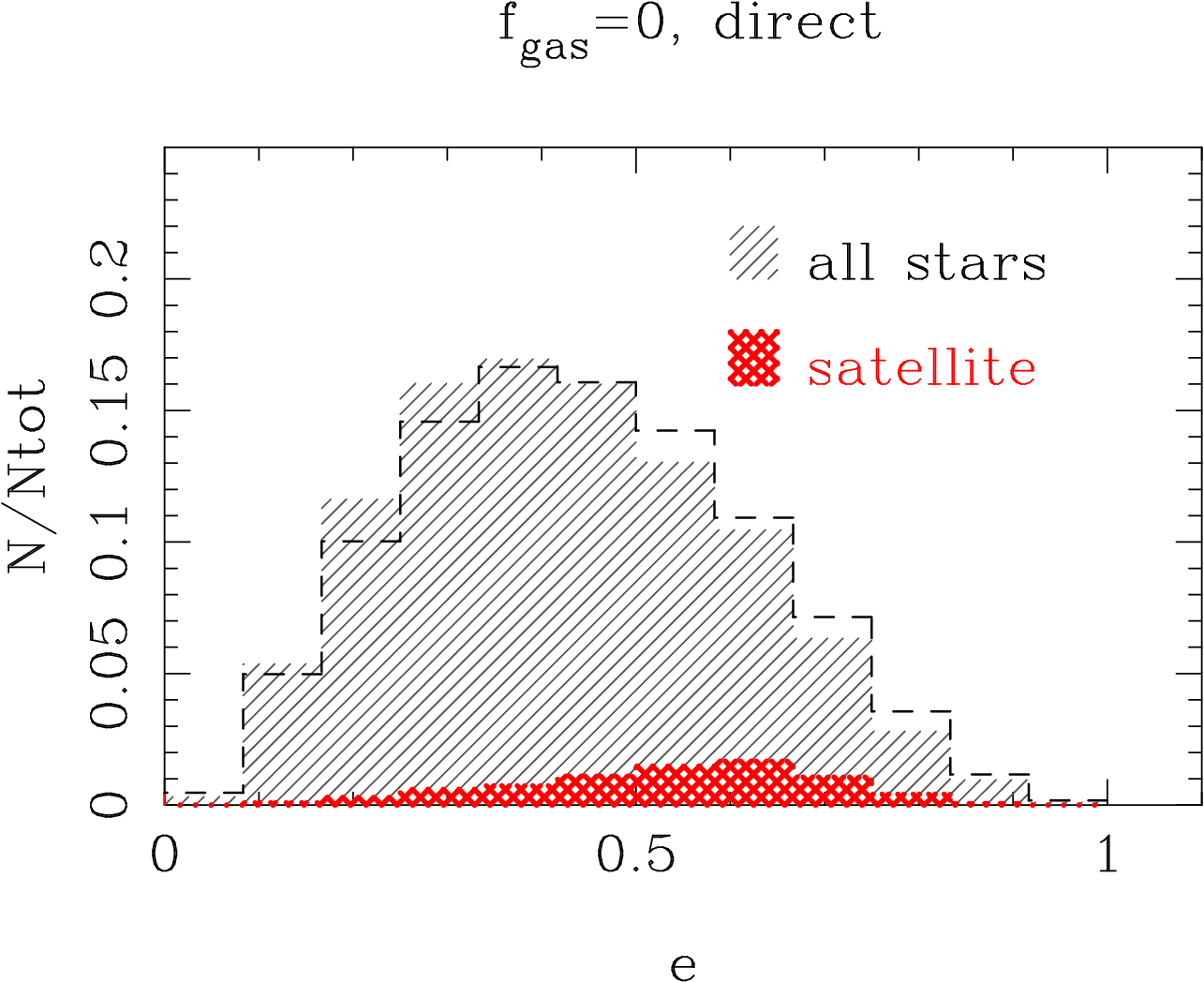}
\includegraphics[width=4.7cm,angle=270]{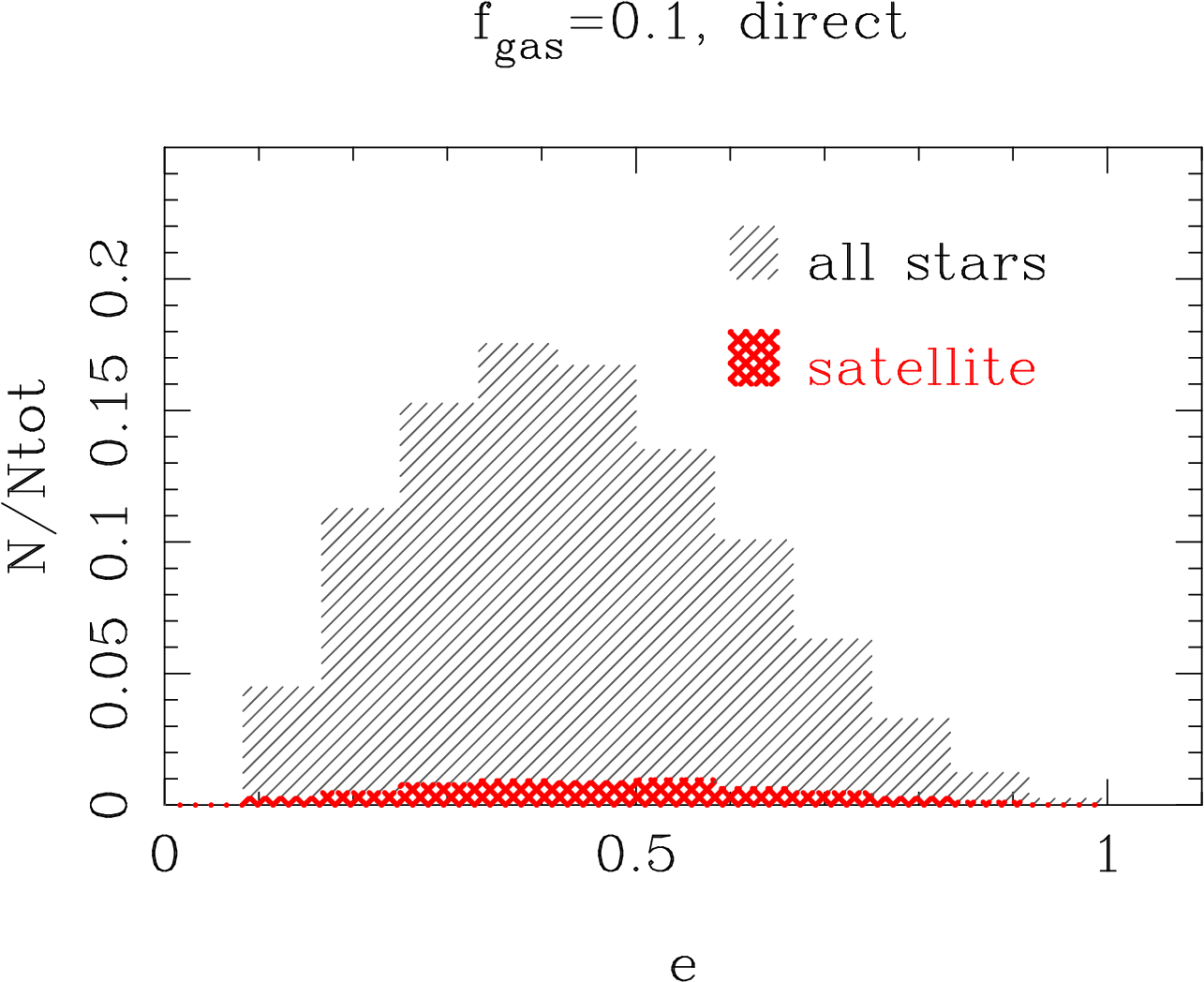}
\includegraphics[width=4.7cm,angle=270]{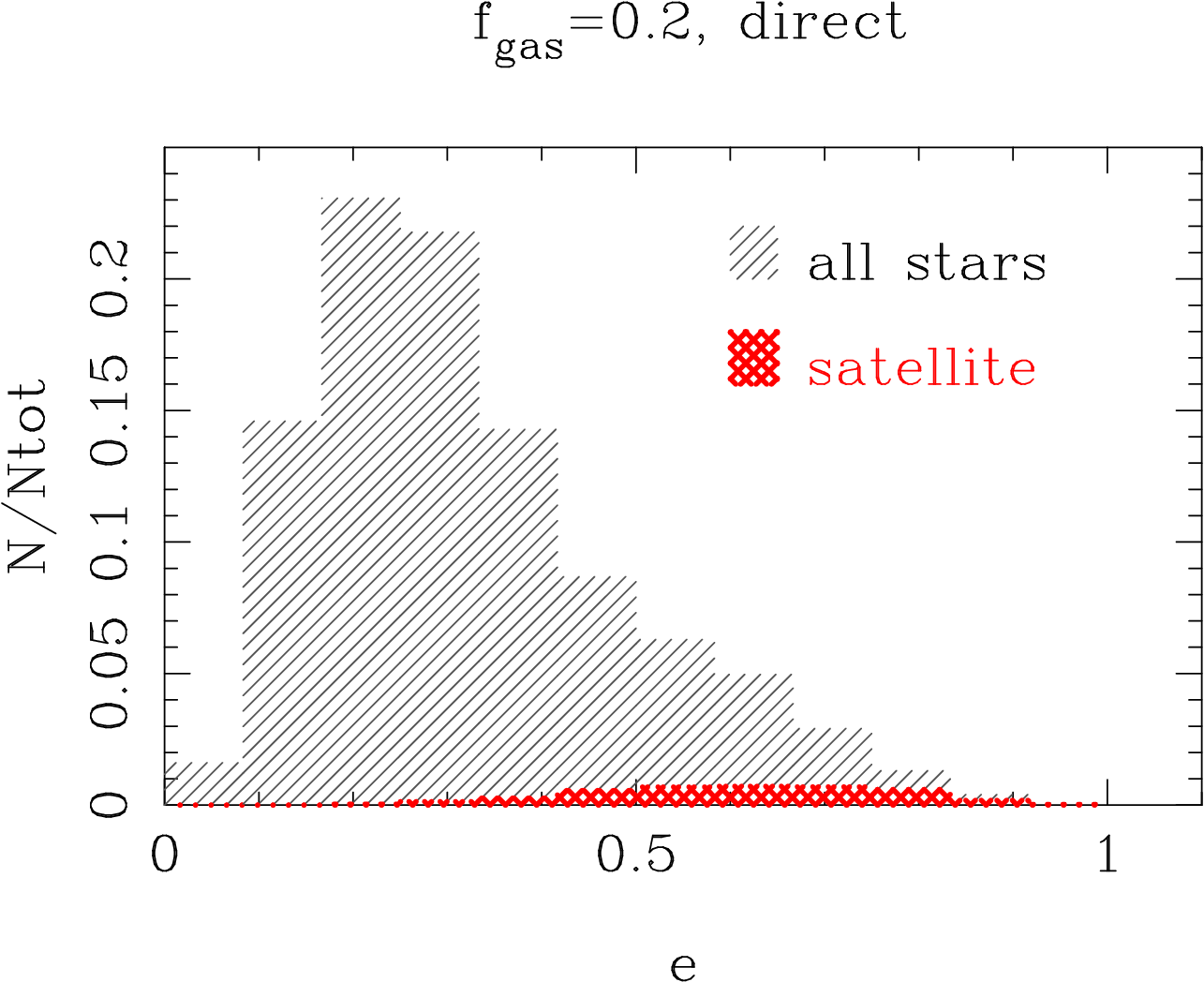}
\includegraphics[width=4.7cm,angle=270]{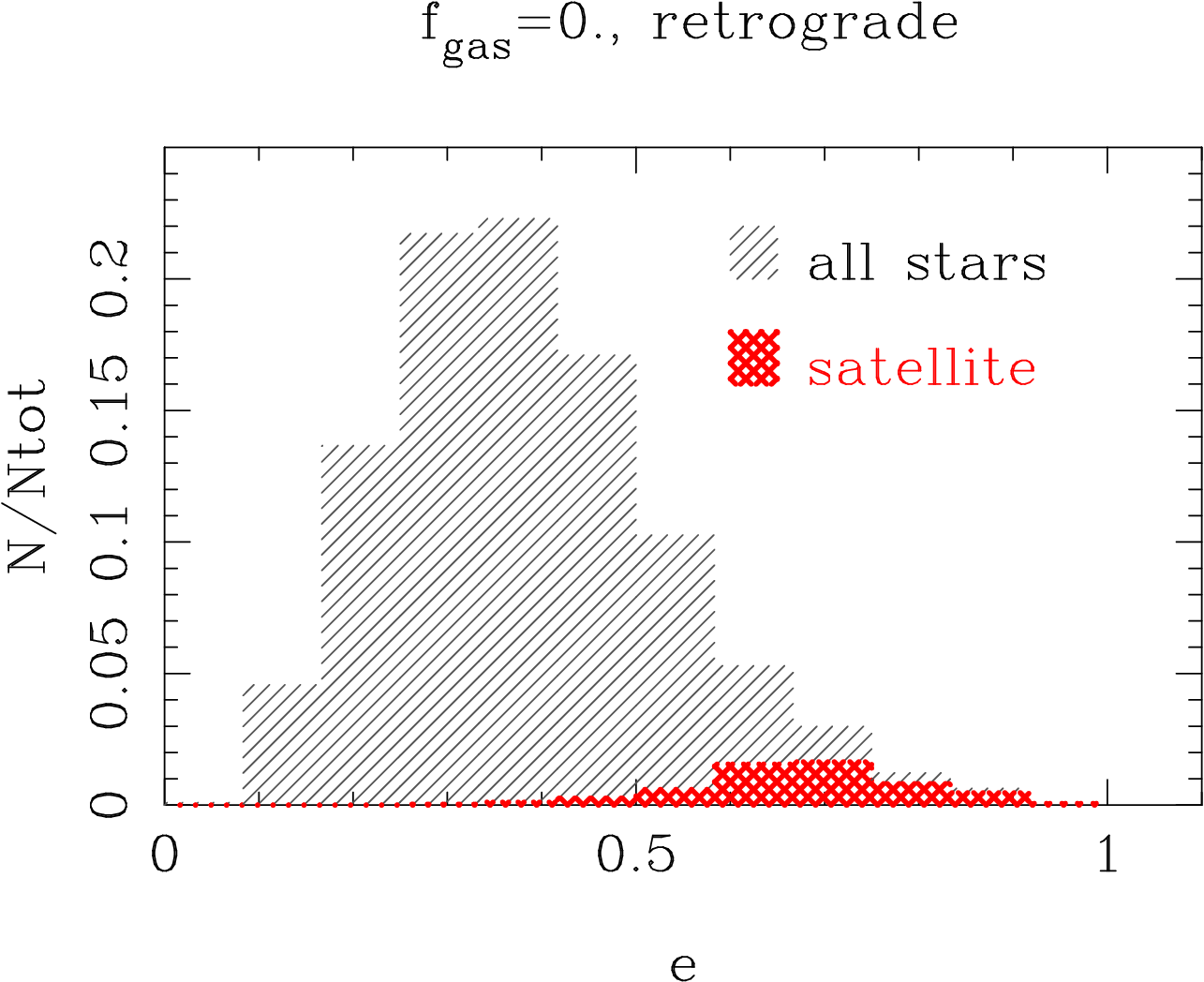}
\includegraphics[width=4.7cm,angle=270]{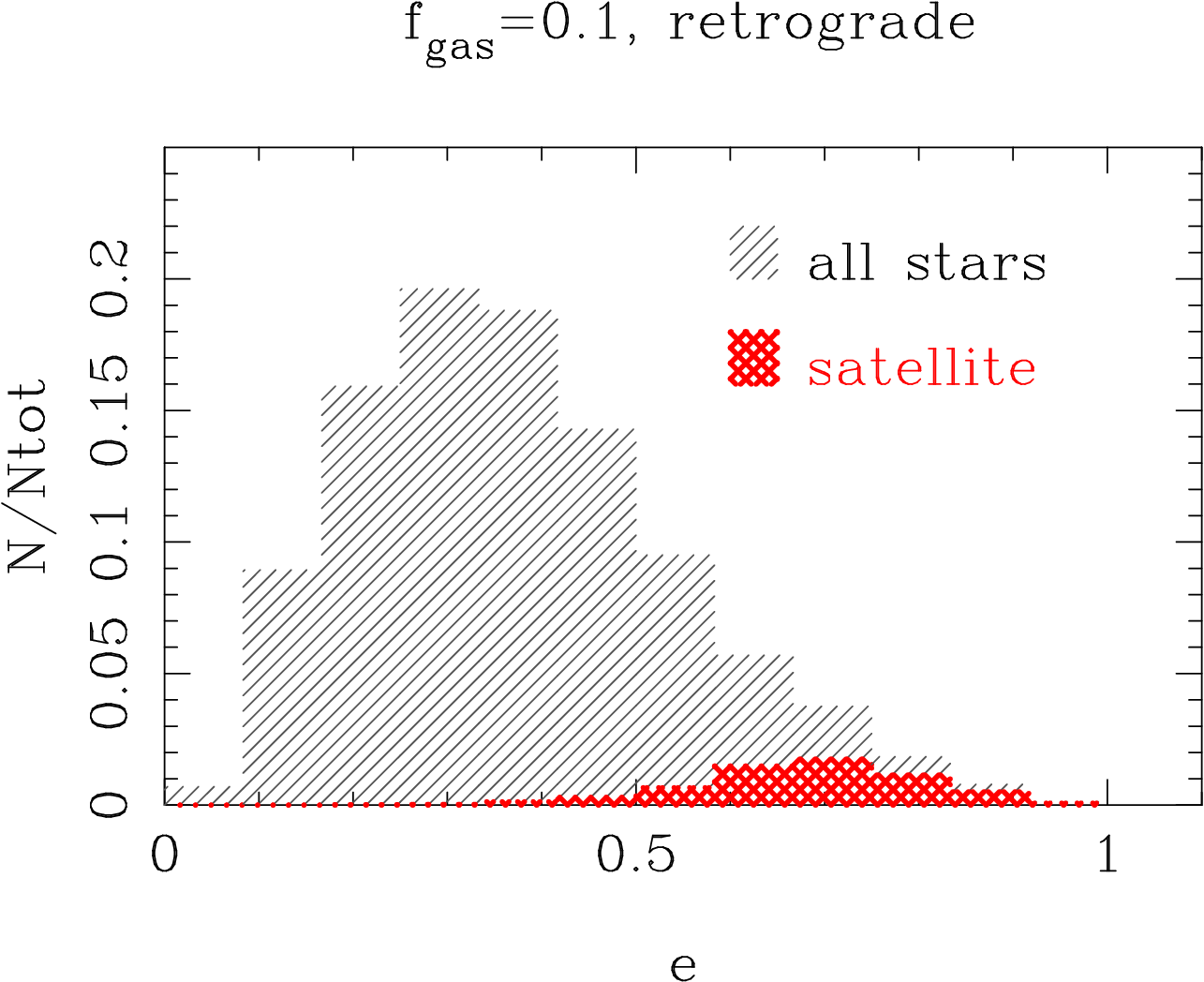}
\includegraphics[width=4.7cm,angle=270]{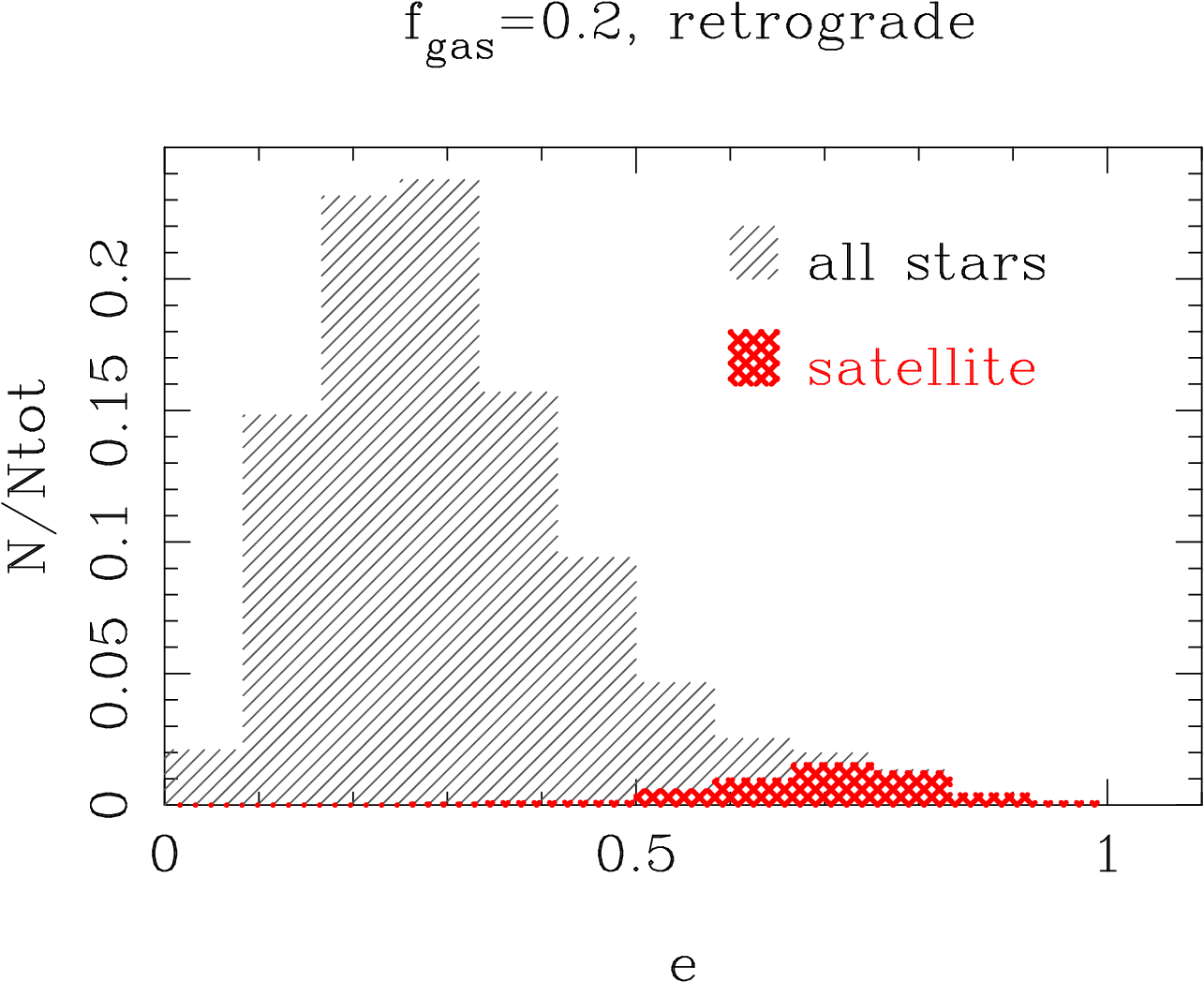}
\vspace{0.7cm}

\caption{Eccentricity distribution of stellar particles in a number of 1:10 minor mergers within a cylindrical
shell at disk scale lengths, $R_d$, from $2 \le R/R_d \le 3$, and at heights, $z_0$, from $1 \le z/z_0 \le 3$.
\emph{From left to right:} The gas-to-stellar disk mass fraction increases from
$f_{gas}$=0, 0.1, and 0.2 in each panel. Mergers on direct orbits are shown in the top row, ones on 
retrograde orbits are shown in the panels of the bottom row. We indicated both the contribution of stars initially in the satellite galaxy (red histogram) and the distribution of all the stars (grey histograms). The dashed histogram in the upper-left panel corresponds to a high-resolution simulation, employing a total of $1~320~000$ particles and a softening length $\epsilon$=150 pc.  The results of the higher resolution simulation are in qualitatively in agreement with those of the lower resolutions simulations.}
\label{all}
\end{figure*}

\section{Results and discussion}\label{results}

We analyzed the resulting eccentricities from the orbits of
the stellar particles in our modeled galaxies during the period 1-2 Gyr after the completion of the merger. This is sufficient for the orbits to be relatively stable. 
 For each particle, we computed the minimal
($r_{min}$) and maximal distance ($r_{max}$) from the galaxy
center during this period of time, and defined the eccentricity
as $e=(r_{max}-r_{min})/(r_{max}+r_{min})$. As done in \citet{sales09}, we restricted our analysis to a
cylindrical region around the galaxy center, at $2\le
R/R_d\le 3$ and $1 \le \mid z/z_0 \mid\le 3$, where $R/R_d$ and $z/z_0$ are the
cylindrical coordinates of a star relative to the disk scale-length and the 
scale height of the thick disk. The disk scale-length varies slightly
from one model to another, depending mostly on the gas fraction in the
primary disk, ranging from $R_d=4.3$ to $R_d=5.4$ kpc.  The average value of the thick disk scale-height in the volume under consideration is  $z_0=2$ kpc (fitted with an exponential), whereas the exact value depends on both the gas fraction of the primary
and the initial orbital configuration \citep{qu2010a}.

For a representative set of models, the resulting distributions of orbital
eccentricities  of the old stars\footnote{By "old stars", we mean
stars already present in the galaxy before the interaction, whereas by
``new'' stars, we mean stars that are formed during the interaction from
the gas, if initially present in the disk. New stars are distributed in both 
a thin disk (scale height of hundreds of parsecs) and the bulge. They
do not contribute to the thick disk.} in the remnant galaxy are shown
in Fig.\ref{all}.  Independent of the exact properties of the galaxies
in the simulations and of the initial orbital configurations, a number
of common features emerge from the simulations:

\begin{itemize}

\item the overall distribution of eccentricities shows a peak at low
values, between $e=0.2-0.35$, followed by a relatively smooth
decrease towards higher values;

\item stars initially in the satellite galaxy span a wide range 
of eccentricities, from $e=0.45$ to $e=0.75$, with the
exact values depending on the orbital characteristics and gas fractions;

\item  we never find a secondary peak at high eccentricities, similar to that between 0.7 and 0.95 reported in \citet{sales09}.

\end{itemize}

Our results are quite different from the distribution found by \citet{sales09}
for minor mergers (labelled ``heating'' in their Figure 3)
-- they resemble more what they found for a thick disk
formed \emph{in situ} by gas-rich mergers early in the evolution of the disk (labelled "merger" in their paper).
What is the cause of this difference from our simulations?

In principle, many parameters can contribute to this difference. The
mass ratio of the encounters is different, for example: we analyze 1:10
mergers, whereas  \citet{sales09} considered a 1:5 ratio. A relatively more massive
satellite would lead to a greater fractional contribution of its stars
to the total mass in the thick disk at a given height above the galaxy
mid-plane. To investigate the impact of a more massive accretion event,
we analyzed the case of a primary galaxy accreting consecutively two
satellites\footnote{Hereafter, we denote simulations of these consecutive
mergers as ``2x(1:10)''.}, each having a mass that is one tenth of the
mass of the primary galaxy. In this case, the total mass acquired through
the two minor mergers is equivalent to that acquired in a single merger
of mass 1:5.  Using the two consecutive mergers with a mass ratio of 1:10
each to investigate the impact of a 1:5 minor merger is appropriate because it
has been shown that the effect of multiple minor mergers is cumulative
\citep{bournaud05, qu2010a}. This being the case, we analyzed several possible consecutive merger
configurations. We considered mergers of two dS0 galaxies
on a gS0,  mergers of two dSa galaxies on a gSa, and mergers of two dSb
galaxies on a gSb; simulating both direct and retrograde encounters.
The result of this analysis is summarized in Fig.2, which represents
the eccentricity distribution of stars that results from a 2x(1:10)
merger of a gSb galaxy and two dSbs on direct orbits. We find that
the consecutive minor mergers cause an increase in the contribution of
satellite stars to the high eccentricity tail of the distribution. We note
also that the overall distribution tends to shift towards higher values
of the eccentricity (this is clear when comparing the results of the direct 1:10 merger
for $f_{gas}=20\%$ in Fig.1 with Fig. 2). In the six cases we analyzed,
we never found the secondary peak identified by Sales et al. 2009. The mass
of the satellite certainly tends to increase the contribution
of satellite stars to the high eccentricity tail of the distribution,
but it appears insufficient in itself to create a second peak. Other
parameters must play a role, perhaps a decisive role in creating
a second peak. For example, the orbital configuration of course has
an effect, but this difference may also depend on the details of the minor merger model analyzed
by Sales et al. (2009) which consists of two galaxies (a primary and
its satellite) embedded in dark halos with a NFW profile. A different
central concentration in the dark matter profile may cause dynamical friction to have a slower/delayed impact allowing
the satellite to survive longer,  thus the orbit to become more
eccentric for a more massive and/or centrally concentrated satellite.
All of this may result in a final eccentricity distribution of stars in
the remnant galaxy that is different from and extends to higher values than those we found.

We do not mean to claim that the distributions of stellar eccentricities
shown in Fig.\ref{all} represent the general outcome for minor mergers,
only that they are a possible outcome, compatible with
 the eccentricity of thick disk stars in the solar neighborhood
\citep{dierickx10, wilson10}.  Nevertheless, the eccentricity distribution
found in our simulations, for the direct merger with a gas fraction of
20$\%$ in the disk, is remarkably similar to those observed in the Milky
Way thick disk. \citet{dierickx10} and \citet{wilson10} found a peak at
low eccentricities with a smooth decline at higher values as we have in
our simulations (Fig.~\ref{comp}).

To  investigate the properties of thick disks formed by minor mergers, we have measured the distribution of
eccentricities of stars at different heights from the disk
mid-plane. This analysis was motivated by observations of changes in the eccentricity
distributions of stars in the solar neighborhood with distance 
from the plane \citep{dierickx10}. We restricted our analysis
to stars in the region between $2 \le R/R_d \le 3$ (Fig.\ref{allvsz}).  We have found that the value of the peak in the overall eccentricity distribution decreases with increasing gas fraction of the primary disk (cf. Fig.\ref{all}
and Fig.\ref{allvsz}). The gravitational force exerted by a gas
component can indeed strongly affect the vertical stellar distribution
\citep{arunima08}, limit disk heating \citep{qu2010a}, and affect the kinematics of the stars in the disk \citep{qu2010b}. 

Interestingly, we also find that the eccentricity distributions as a
function of $z$ show a dependency on the initial orbit of the satellite
galaxy. If the satellite is accreted along a direct orbit (i.e. if the
$z$ component of the orbital angular momentum is parallel to the primary
internal angular momentum), the eccentricities show a trend with the vertical 
distance from the mid-plane: \emph{the larger the distance from 
the mid-plane, the greater the typical stellar eccentricity}. This trend can be
explained in terms of a rotational lag of the thick disk as a function
of $z$, which can be induced by minor mergers \citep{villalobos08}
but whose amplitude and behavior strongly depend on the satellite
orbit \citep{qu2010b}. In direct encounters, the strong coupling of
the satellite motion to the disk internal rotation induces a higher
perturbation of the kinematics of stars in the primary disk, increasing
the overall radial velocity dispersion and decreasing the overall
rotational support. This effect also manifests itself in the amplitude of the
rotational lag of the thick disk.  As the stars in the thick disk mostly
originated in the thin disk of the primary, the lag must increase with
$z$. For retrograde orbits, however, the lack of strong gravitational coupling
between satellite and disk produces a much smaller effect -- stars in
the thick disk rotate slower than those in the thin disk, but this lag
does not show any dependence on vertical height, except at very large
mid-plane distances (above $z > 5z_0$). The absence of a trend between the
rotational support and $z$ for retrograde encounters is reflected directly
in the distribution of eccentricities, which does not change significantly
with increasing $z$. In particular, the peak of the distribution is
constant at $\mid z/z_0 \mid \le$5, and increases only at even larger disk heights.

Thus the observational trend found by \citet{dierickx10}, where stellar orbits become
more eccentric as one moves away from the Galactic plane, can only be reproduced
by our models if the satellite is accreted along a direct orbit.

\begin{figure}
\centering
\includegraphics[width=5.5cm,angle=270]{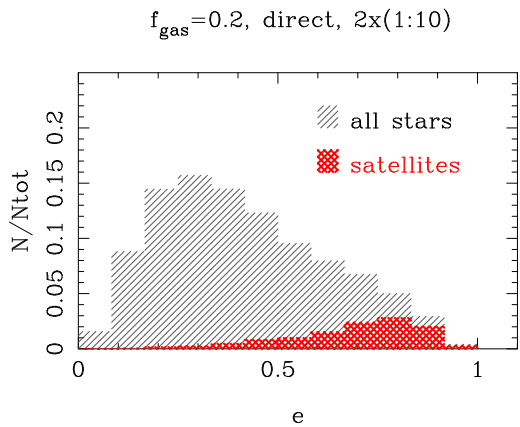}
\vspace{0.7cm}
\caption{Eccentricity distribution of thick disk stellar
particles for a  2x(1:10) consecutive minor merger with a gas fraction
$f_{gas}$=0.2. As in Fig.~1, stars have been selected in a cylindrical
shell at disk scale lengths, $R_d$,  from $2 \le R/R_d \le 3$, and at
heights, $z_0$, from $1 \le z/z_0 \le 3$.  The contribution of stars
accreted from the satellite galaxies is shown in red.}
\label{higher}
\end{figure}

\begin{figure}
\centering
\includegraphics[width=5.cm,angle=270]{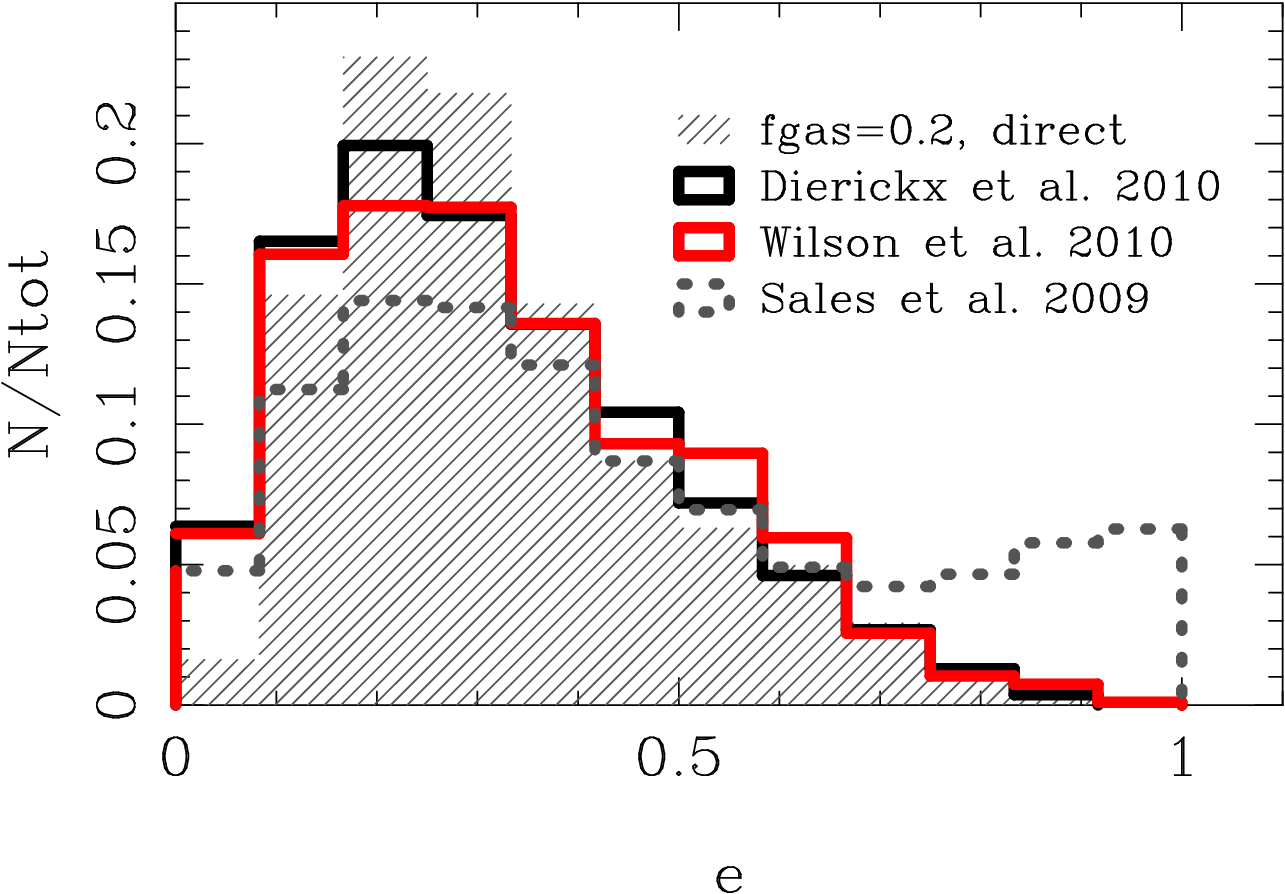}
\vspace{0.7cm}
\caption{Comparison of the eccentricity distributions of stellar
particles of one of our minor merger models with the distributions found by \citet{dierickx10} and \citet{wilson10} for stars
in the solar neighborhood. For comparison, the distribution found by \citet{sales09}
for their ``heating'' (i.e., minor merger) scenario is also shown. In particular, one should note the secondary peak at high values of $e$, which is not seen in our simulations of 1:10 minor mergers.}
\label{comp}
\end{figure}

\begin{figure}
\centering
\includegraphics[width=5.cm,angle=270]{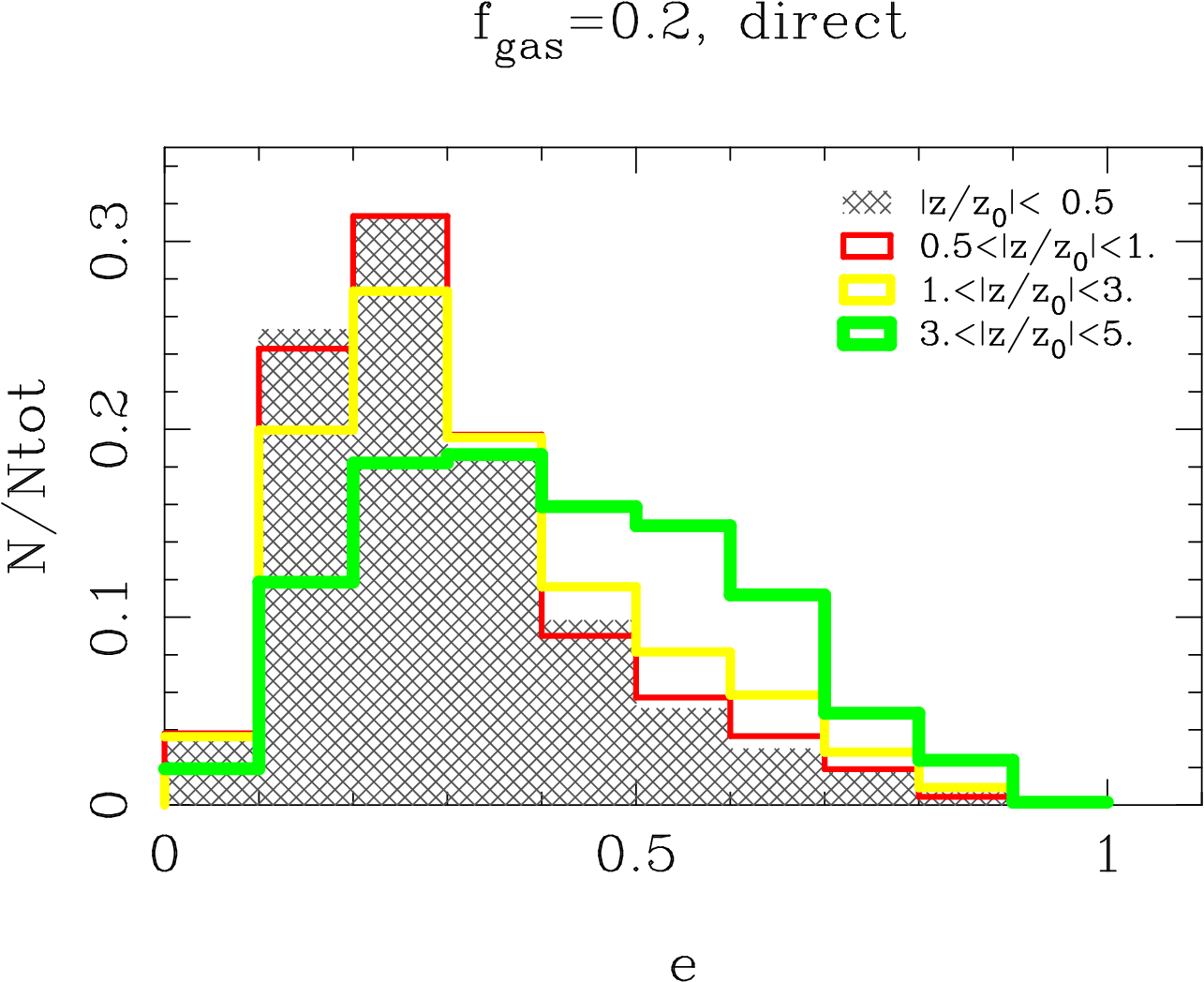}
\includegraphics[width=5.cm,angle=270]{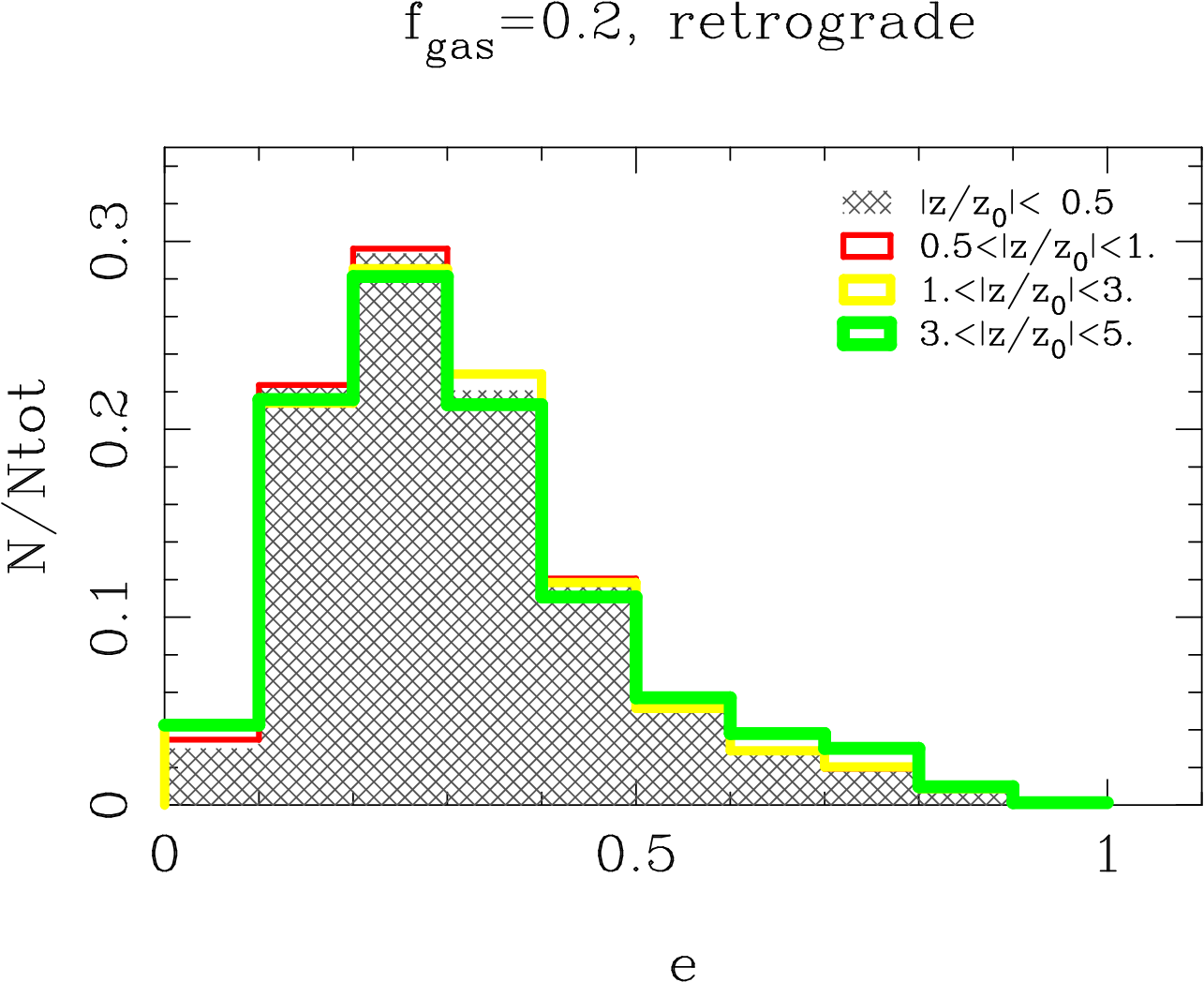}
\vspace{0.7cm}
\caption{Height-dependence of the stellar eccentricity distribution in a cylindrical
shell with $2 \le R/R_d \le 3$, for a direct (top row) and a retrograde (bottom row) 1:10 merger, with $f_{gas}=20\%$ . In each
panel, the histograms correspond to different heights $z$ above and below the
galaxy mid-plane as indicated in the legend.}
\label{allvsz}
\end{figure}


\section{Conclusions}\label{conclusions}

By means of N-body/SPH simulations, we have investigated the distribution
of orbital eccentricities of stars in thick disks formed by heating
of a thin stellar disk after minor mergers with 1:10 mass ratios. The
simulated interactions span a range in both gas fractions of the primary disk
and initial orbital parameters. The eccentricity distributions have
an approximately triangular shape, with a peak around $e=0.2-0.35$, and
a relatively smooth decline towards higher eccentricities.  The exact
distribution depends on the gas-to-stellar disk mass fraction in the primary disk, in the sense that higher gas fractions lead to lower peak eccentricity values. It also depends on the orbital
configuration -- direct orbits tend to produce broader eccentricity distributions than retrograde encounters.  Stars that originated from
the satellite galaxy tend to have relatively high eccentricities after
the merger is completed (on average $e$=0.45 to 0.75), similar to those recently found in simulations by Sales
et al. (2009) but do not reach the highest values that these authors found nor show their secondary peak at  $e \sim 0.8$.
This is likely a consequence of the lower satellite mass of the satellites in our
simulations, of the different dark-matter halo profiles used to model the
galaxies, and our greater range of initial orbital configurations.
The absence of this secondary peak at high eccentricities, in all our models, results in a
distribution of eccentricities that matches qualitatively the observations \citep{dierickx10, wilson10}.  Moreover, the increase
in the orbital eccentricities of stars in the solar neighborhood
with vertical distance from the galaxy mid-plane found by Diericxk et
al. (2010) can also be qualitatively reproduced by our models, but only if
the merging satellite is accreted on a direct orbit.  No trend is
found for retrograde orbits, except for stars high above the mid-plane
at distances greater than $5 z_0$.

On the basis of these models, we conclude that the observed distribution
of orbital eccentricities of stars in the Galactic thick disk is in
qualitative agreement with a scenario where the thick disk has been
(predominantly?) formed by heating the thin disk during the impact of a
(or perhaps multiple) minor merger(s) \citep[whose effects are likely
cumulative][]{qu2010a}. As we have discussed, the formation of a thick
disk {\it in situ}, i.e., after the merger of gas-rich progenitors
in the early universe, cannot be ruled out based on our models or
other evidence, because its observed eccentricity distribution is also
qualitatively similar to that observed.

For a satellite accreted on a direct orbit, our models can also
reproduce the trend found by \citet{dierickx10} that satellite orbits
become more eccentric as one moves away from the Milky Way plane.
We speculate that if minor mergers were the dominant means of
formating the Milky Way thick disk, the major contributor should have been 
 merger(s) with satellite(s) on direct orbits.  If the thick
disk of the Milky Way was formed by external heating, other processes
may have contributed to its subsequent evolution and helped to shape its
current characteristics. Radial migration  \citep{selbin02, roskar08, roskar08b}, for example, is particularly
efficient if asymmetries such as bars and spiral arms develop in the thin
disk \citep{debattista06, foyle08, minchev10a, minchev10b} and may well have played an
important role in influencing the kinematic and chemical properties
of the early formed thick disk \citep{loebman10}. Crucial tests
of the viability of processes such as minor mergers will be made through
 detailed comparisons that are not restricted to eccentricity data but will involve the kinematic properties and metal abundances
and ratios of the stars of the thick disk.

The presently available computational facilities ensure that the investigation of a large
number of parameters and accretion histories is feasible. By comparing the
signatures produced by different accretion events in the kinematics and
metallicity of the thin/thick disk and halo of the Galaxy, the future observations by GAIA
and LAMOST, for example, will have a key role in understanding the mass
accretion history of our Galaxy and its impact on the formation and
evolution of the thick disk.

\section*{Acknowledgments}
YQ and PDM are supported by a grant from the French Agence Nationale de
la Recherche (ANR). We are grateful to Beno\^it Semelin and Fran\c{c}oise
Combes for developing the code used in this paper and for their permission
to use it.  These simulations will be made available as part of the
GalMer simulation data base (\emph{http://galmer.obspm.fr}).
We thank the referee for her/his comments and suggestions, which improved the clarity of the paper and the presentation of the results.

\end{document}